# MultiCaM-Vis: Visual Exploration of Multi-Classification Model with High Number of Classes

Syed Ahsan Ali Dilawer*  
University of Kaiserslautern, Germany.

Shah Rukh Humayoun†  
San Francisco State University, USA.

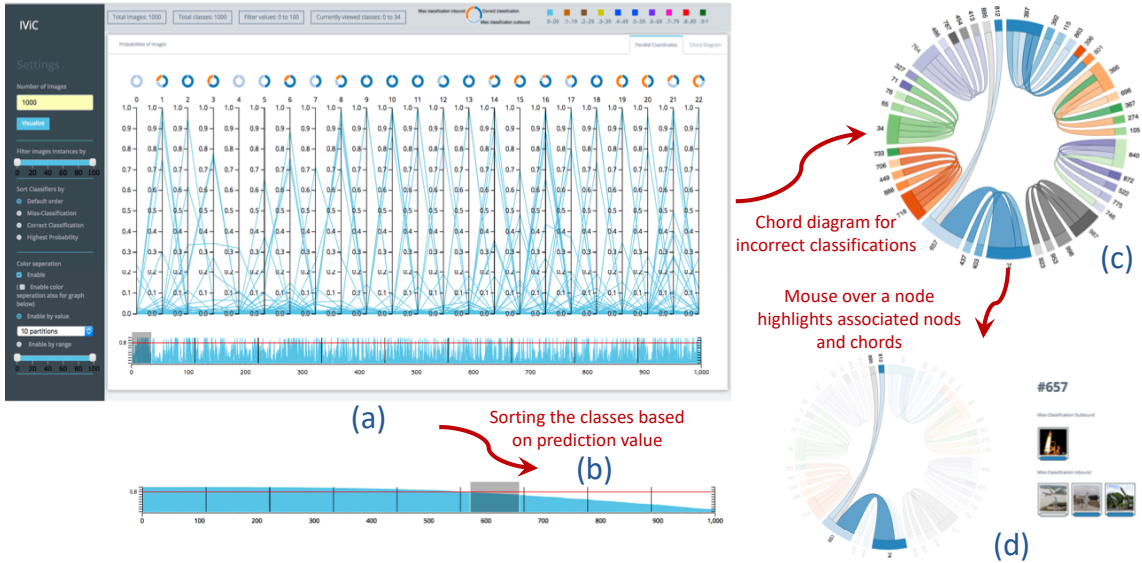

Figure 1: **(a)** MultiCaM-Vis model exploration view using two parallel coordinates in *Overview+detail* style. **(b)** Detailed view sorted based on prediction value. **(c)** Chord diagram view to inspect incorrect classification cases. **(d)** Mouse hover a particular node (class) in (c) highlights the associated chords and nodes as well as images of associated incorrect classification cases.

**ABSTRACT**

Visual exploration of multi-classification models with large number of classes would help machine learning experts in identifying the root cause of a problem that occurs during learning phase such as miss-classification of instances. Most of the previous visual analytics solutions targeted only a few classes. In this paper, we present our interactive visual analytics tool, called **MultiCaM-Vis**, that provides overview+detail style parallel coordinate views and a Chord diagram for exploration and inspection of class-level miss-classification of instances. We also present results of a preliminary user study with 12 participants.

## 1 INTRODUCTION

Convolution Neural Network (CNN) approach in Machine Learning (ML) uses combinations of convectional layers and pooling layers to reduce the computational cost and output variations. therefore, it has been applied in different domains from recognizing letters to identify objects in images [3]. However, these all layers typically reside in a black-box, which makes it difficult for ML experts to identify the root cause of problems that may occur during the learning phase such as misclassification of instances. The traditional backtracking solutions are time consuming and hectic for ML practitioners.

*e-mail: ahsan.shah777@gmail.com  
†e-mail: humayoun@sfsu.edu

Targeting this concern, many visual analytics tools have been proposed targeting multi-classification models. Few examples are: Seifert & Lex [8] targeted classification models with 20 distinct classes, whose predictions can be interpreted as probabilities at three levels, i.e., classifier level, class level, and test item level. Paiva et al. [4] developed a point-based visualization system for maximum 10 classes. Alsallakh et al. [1] developed a diagnostic VA system to depict the prediction score distributions, which was efficient for lower than 20 classes. Ren et al. [6] developed Squares tool to support performance diagnosis within a single visualization while showing prediction score distributions at multiple levels of detail, useful for less than 20 classes. Recently, Pommé et al. [5] utilized confusion matrices and color encoding to display the class-wise differences of performances between two classification models having up to 30 classes, while Uwaseki et al. [9] targeted multi-class image classification models having less than 10 classes.

From the above-mentioned work, it can be concluded that visual exploration of multi-classification models with large number of classes are not adequately addressed in the literature. Targeting this concern, we present an interactive VA tool, called **MultiCaM-Vis**: **Multi-Cla**ssification Model **Vis**ualizer, that allows users to visual explore multi-classification model, targeting 1K classes in ILSVRC image dataset [7], using parallel coordinate views in *overview+detail* style. Users can explore the misclassification (inbound and outbound) cases to get better understanding of problematic classes in the model. The MultiCaM-Vis also provides a Chord diagram view for in-depth inspection of the incorrect classification cases. Several filtering and sorting options are provided at each level to help users in model exploration and inspection. We also conducted a preliminary user study with 12 participants to check the effectiveness of the tool as well as to get their feedback.

## 2 THE DATASET

ImageNet [2] dataset consists of a collection of over 15 million high resolution labelled images. These images belong to an estimated 22k different categories. In our tool, we use ImageNet Large-Scaled Visual Recognition Challenge (ILSVRC) dataset [7], a subset of ImageNet, with images belong to 1k different categories. Overall, ILSVRC contains an estimated 1.2 million training images, 50k validation images and 150k testing images. From ILSVRC subset, we use set of validation images with 500 X 500 color photograph in 1K different object classes. The used network in our case consisted of eight learned layers, five convolutional layers and three fully connected layers. The output of the last fully connected layer (fc8) is a prediction distribution score over the 1000 class labels.

## 3 THE MULTICAM-VIS TOOL

Our developed tool, called **Circles**, visualizes results of a multi-classification model targeting 1K classes in ILSVRC [7] dataset. The web-based client side was developed using HTML, CSS, JavaScript, and D3.js library while the server side was developed using Node.js.

In order to explore a model's results, MultiCaM-Vis provides two parallel coordinates using *overview+detail* style to show the class-level prediction distribution across all 1K image classes in ILSVRC dataset (Fig. 1(a)). The bottom *overview* parallel coordinate view shows all the 1K classes with class-level prediction distribution, while the upper parallel coordinate *detailed* view initially shows the first 10 classes. MultiCaM-Vis provides an interactive slider on the overview view, where users can select the range of classes to be shown on the detailed view. On the top of each class in the detailed view, MultiCaM-Vis provides a doughnut chart to show the distribution of *correct-classification* instances in dark blue color, the *incorrect-classification-inbound* instances in light blue color, and *incorrect-classification-outbound* instances in orange color.

MultiCaM-Vis provides a number of filtering and sorting options. For example, mouse hover a particular image instance line in detailed view shows the associated image with the species name and its hierarchy in the ILSVRC dataset. MultiCaM-Vis provides a filtering and sorting panel on the left-side with a number of options (left-side in Fig. 1(a)). Few examples are: a slider-range to show only those image instances that have a corresponding prediction value within this slider range against one or more classes; sorting of the classes in the both parallel coordinates views order by incorrect-classification (in-bound or out-bound) cases, order by correct-classification cases, or order by instances' highest probability distribution; giving different colors to image instances based on prediction distribution where users can have the option of the range from 1 to 10 colors (in this case, if a user selects 10 colors option, then MultiCaM-Vis associates a color to an instance based on its highest prediction to a class where each color range is 0.1 of total 1 prediction probability) or by a range-slider where instances with having a prediction probability within this range will get a different color than all other instances.

To provide in-depth inspection of incorrect-classification (*inbound* or *outbound*) cases, MultiCaM-Vis opens another view using a Chord diagram based on demand. In this case, MultiCaM-Vis shows only those classes in the Chord diagram that are selected by users from the detailed view (Fig. 1(c)). The chords in this view represent incorrect-classification cases. MultiCaM-Vis allocates color to each chord as per the correct class (node) color. The width of a chord represents the number of incorrect-classification cases from one class to another class (inbound or outbound). Mouse hover a particular chord shows the associated image instances with correct classes while mouse hover a particular class (node) highlights only this class (node) and the associated classes (nodes) and chords as well as all the image instances with inbound and outbound incorrect-classification cases ((Fig. 1(d)).

## 4 THE PRELIMINARY USER STUDY

We conducted a preliminary user study with 12 participants (4 females, 7 had ML background), all graduate students (age range: 25 to 30, M = 28.3). We were interested to see the effectiveness of the tool in term of *accuracy* and to know whether participants from ML and non-ML background have any differences in accuracy. We designed 3 controlled tasks, with subtasks, for the study in a way that we could acquire the reaction about participants' interaction with the tool and to identify whether they are able to find appropriate information or not. The first task targeted towards finding out correct or incorrect classification cases in both parallel coordinates and the Chord diagram, the second task targeted towards using different filtering and sorting options, while the third task targeted the use of color options in the resulting parallel coordinates. The experiment sequence was: filling the demographic form, tool tutorial, tasks execution, a 10 closed-ended questionnaire, and finally getting feedback in general. Each experiment lasted no more than 1 hour.

In the case of accuracy, ML participants achieved an overall 94% accuracy in all three tasks compared to 85.67% for non-ML participants (i.e.: **T1** ML: 91%, non-ML: 82%; **T2** ML: 98%, non-ML: 89%; **T3** ML: 93%, non-ML: 86%). These results are quite promising as it shows that even non-ML background participants were quite good in accurately executing all three tasks with little difference compared to ML participants. Table 1 shows participants' feedback in the closed-ended questionnaire. In the future, we plan to add the inter-model comparison view so users can compare between multiple models of the same dataset. Also, we intend to execute a detailed user study targeting the new inter-model comparison view.

| | Question | 1 | 2 | 3 | 4 | 5 | M |
|---|---|---|---|---|---|---|---|
| 1. | *Intuitiveness* of visualization | 0 | 0 | 2 | 7 | 3 | 4.08 |
| 2. | *Efficiency* in completing tasks | 0 | 0 | 2 | 6 | 4 | 4.17 |
| 3. | *Clarity* of visualization in finding the required information | 0 | 0 | 1 | 3 | 8 | 4.58 |
| 4. | *Support the analysis* of correct and in-correct classification | 0 | 0 | 0 | 4 | 8 | 4.67 |
| 5. | *Easy and intuitiveness* of filtering and navigation | 0 | 0 | 3 | 5 | 4 | 4.08 |
| 6. | *Usefulness* of visualization in finding classification issues | 0 | 0 | 0 | 10 | 2 | 4.16 |
| 7. | *Easy to use* tool and visualizations | 0 | 0 | 4 | 6 | 2 | 3.83 |
| 8. | *Easy to learn* tool and visualizations | 0 | 0 | 2 | 6 | 4 | 4.16 |
| 9. | *Future usage* of the tool | 0 | 0 | 6 | 4 | 2 | 3.67 |
| 10. | *Recommending* the tool to others | 0 | 0 | 0 | 3 | 9 | 4.75 |

Table 1: Participants' feedback score in closed-ended questionnaire